\documentclass{aa}  
\usepackage{graphicx}
\usepackage{txfonts}
\begin{document}
\title{Contamination by field late-M, L and T dwarfs in~deep~surveys} 
\titlerunning{Field late-M, L and T dwarfs in deep surveys}
%
%
\author{J.~A. Caballero\inst{1,2}
    	\and
     	A. J. Burgasser\inst{3}
    	\and
     	R. Klement\inst{1}     
}
\offprints{Jos\'e Antonio Caballero, \email{caballero@astrax.fis.ucm.es}}
\institute{
Max-Planck-Institut f\"ur Astronomie, K\"onigstuhl 17, D-69117 Heidelberg,
Germany 
\and 
Dpto. de Astrof\'{\i}sica y Ciencias de la Atm\'osfera, Facultad de
Ciencias F\'{\i}sicas, Universidad Complutense de Madrid, E-28040 Madrid, Spain
\and 
Massachusetts Institute of Technology, Kavli Institute for Astrophysics and
Space Research, Cambridge, MA 02139, USA 
}
\date{Received 6 February 2008 / Accepted 26 May 2008}

\abstract
{Deep photometric surveys for substellar objects in young clusters and for
high-redshift quasars are affected by contaminant sources at different
heliocentric distances.  
If not correctly taken into account, the contamination may have a strong effect
on the Initial Mass Function determination and on the identification of
quasars.}   
{We calculate in detail the back- and foreground contamination by field dwarfs
of very late spectral types (intermediate and late M, L and T) in deep surveys
and provide the data and tools for the computation.}
{Up-to-date models and data from the literature have been used:
($i$) a model of the Galactic thin disc by an exponential law;
($ii$) the length and height scales for late-type dwarfs;
($iii$) the local spatial densities, absolute magnitudes and colours of dwarfs
for each spectral type.}  
{We derive a simplified expression for the spatial density in the thin
disc that depends on the heliocentric distance and the galactic coordinates
($l,~b$) and integrate it in the truncated cone screened in the survey.
As a practical application, we compute the numbers of L- and T-type field
dwarfs in very deep ($I$ = 21--29\,mag) surveys in the direction of the young
$\sigma$~Orionis cluster.
The increasing number of contaminants at the faintest magnitudes could
inhibit the study of the opacity mass limit at $M \lesssim$ 0.003\,$M_\odot$
in the~cluster.} 
{}
\keywords{stars: low mass, brown dwarfs -- stars: luminosity function, mass
function -- Galaxy: stellar content -- open clusters and associations:
individual: $\sigma$~Orionis -- methods: analytical}   
\maketitle
%

\section{Introduction}
\label{introduction}

Deep wide-field multi-band photometric searches in open clusters and
star-forming regions are the most useful tool for the discovery and
identification of young brown dwarfs and planetary-mass objects (e.g., Rebolo
et~al. 1996 and Bouvier et~al. 1998 in the \object{Pleiades}; Preibisch \&
Zinnecker 1999 and Ardila, Mart\'{\i}n \& Basri 2000 in \object{Upper
Scorpius}; Zapatero Osorio et~al. 2000 and B\'ejar et~al. 2001 in
\object{$\sigma$~Orionis}).
Unfortunately, an important fraction of the selected cluster member candidates
do not belong to the cluster, but actually are interlopers in the fore- and
background. 
This fact strongly affects posterior analyses of the bottom of the Initial Mass
Function, down to about a few Jupiter masses. 
There are several ways of reducing the contamination of the mass function
incurred by these interlopers, including the identification of spectral
indicators of youth/low surface gravity in individual sources (Mart\'{\i}n,
Rebolo \& Zapatero Osorio 1996; Stauffer, Schultz \& Kirkpatrick 1998; White \&
Basri 2003; McGovern et~al. 2004; Mohanty, Jayawardhana \& Basri 2005) and
removing sources with kinematics inconsistent with the cluster (Stauffer,
Hamilton \& Probst 1994; Hambly et~al. 1999; Moraux, Bouvier \& Stauffer 2001;
Bihain et~al. 2006; Lodieu et~al. 2007b).  
However, these techniques may require prohibitive investments of time on large 
telescopes, particularly for the lowest mass objects (e.g., S\,Ori~70 -- see
below), and may not eliminate contaminants for distant and/or older clusters and
associations.  
For example, at 100\,pc, accuracies better than $\sim$10\,mas/\,a$^{-1}$ are
needed to measure velocities of $\sim$5\,km\,s$^{-1}$. 
In that case, there can still be foreground objects that share the same motions
but are {\em not} cluster members.
Their membership status can only be insured through the detection of youth
signatures. 
Some known features of extreme youth (age $\lesssim$ 10\,Ma) are Li~{\sc
i} $\lambda$6707.8\,\AA~in absorption (Rebolo, Mart\'{\i}n \& Magazz\`u
1992), abnormal strength of alkali lines and water and H$_2$ absorption bands
due to low gravity (Mart\'{\i}n et~al. 1999; Kirkpatrick et~al. 2006; Allers
et~al. 2007 -- H$_2$ can also be in emission from discs [Thi et~al.
2001]), strong H$\alpha$ $\lambda$6562.8\,\AA~emission associated to 
accretion and mid-infrared flux excesses due to circum(sub)stellar~discs. 
However, obtaining high signal-to-noise ratio, mid-resolution spectroscopy or
{\em Spitzer} Space Telescope imaging of the faintest, reddest objects in a deep
photometric survey ($I \gtrsim$ 18--19\,mag, $J \gtrsim$ 15--16\,mag) can be a
very difficult~task.

A useful statistical approach is to measure both the luminosity function in the
area of the cluster and in neighbouring fields, and extracting the true
cluster luminosity function as the difference of these. 
This method has been succesfully employed in the \object{Orion Nebula Cluster}
and described in detail by, e.g., Muench et~al. (2002). 
In addition to relying on a local extinction map of the cluster to redden the
stellar population in the extinction-free comparison fields, the method 
has, regrettably, an important handicap: 
one must integrate the same exposure time in the neighbouring fields to properly
subtract the ``field mass function''.
For the deepest surveys in clusters, that use expensive large facilities, this
time-consuming solution cannot be accomplished in many cases.
As a result, the vast majority of these deep surveys establish certain selection
criteria for cluster membership based on colour-magnitude diagrams.

Depending on the criteria and the number and suitability of passbands, more or
less cluster non-member contaminants could be among the selected member
candidates in a photometric survey.
To define a selection criterion, the authors must account for photometric
uncertainties in the survey, the natural scatter of the cluster sequence (i.e.,
the sources themselves do not have a single absolute magnitude or colour, but
rather a range of both appropriate for their spectral type), unresolved binarity
and intrinsic photometric variability. 
Figure~2 in Caballero et~al. (2007) illustrates a typical selection of member
candidates in the $\sigma$~Orionis cluster based on the loci of
spectroscopically confirmed members in an $I$ vs. $I-J$ diagram.
Theoretical isochrones may be used as reference, as well.
The removal of interlopers, whose location coincides with that of cluster
members in colour-magnitude diagrams, is very important in the faintest and
reddest magnitude intervals.
Quantitative estimations of contaminants have been presented in several searches
in young clusters (Paresce, de Marchi \& Romaniello 1995; Zapatero Osorio,
Rebolo \& Mart\'{\i}n 1997; Mart\'{\i}n et~al. 2000; Lucas et~al. 2001; Jeffries
et~al.~2004; Moraux et~al.~2007). 

In this paper we provide the necessary tools and state-of-the-art data for the
correct decontamination of field late-type stars and brown dwarfs in deep
photometric surveys at intermediate and high galactic latitudes. 
A practical application of the integrated number of contaminants in very deep
surveys towards the $\sigma$~Orionis cluster is presented.
Our work can be applied to searches in other open clusters and star-forming
regions, {\em but also to high-$z$ quasar surveys.}

\section{Analysis}

\subsection{Possible contaminants}
\label{possible}

The contaminants appearing in deep photometric searches for very red
objects are in general faint reddened galaxies, variable subgiants,
(reddened) distant giants, or field dwarfs with very late spectral types
that fall close to the sequence of the cluster under study in a colour-magnitude
diagram:  
\begin{itemize}  

\item Galaxies are in general extended sources, so their FWHMs are larger than
the average FWHM of the stars in the studied field of view, and they are easily
rejected during the point-spread-function photometry:
down to $J \sim$ 19--20\,mag, galaxies can be revealed under good seeing
conditions (see, for example, the recent work by Foster et~al. [2008]).
They also display colours that do not match the dwarf sequence in a
colour-colour diagram (e.g., $J-K_{\rm s}$ vs. $I-J$). 
Some remote quasars with $z \sim 6$ have, however, point-like FWHMs and spectral
energy distributions nearly identical of those of mid-T dwarfs; these quasars
are, nevertheless, extremely rare (Stern et~al.~2007).

\item Except for extremely young star forming regions with high extinction
clouds (\object{$\rho$~Ophiuchi}, \object{Chamaeleon~I}, Orion
Nebula Cluster), open clusters are relatively free of interstellar absorption.
In some clusters, as $\sigma$~Orionis, the strong wind and ultraviolet radiation
from the Trapezium-like central system has completely cleared the region from
cometary globulae, molecular clouds and most of the original intracluster
material. 
In the case of clusters with thick molecular clouds, these can act as a ``shield
to background field stars'' (Lucas \& Roche 2000; Luhman et~al. 2000; Muench et
al. 2002), which may facilitate the decontamination (there will be only
interloping non-cluster field stars in~foreground).

\item According to Kirkpatrick et~al. (1994), contamination by giant stars in
areas excluding the Galactic plane is negligible.
Given the absolute magnitude of late-type giant stars, a hypothetical faint
(intrinsically bright) giant contaminant would be at several kiloparsecs
over or below the Galactic plane, where only the globular clusters are found. 
More recent works have shown a comparatively large number of giant interlopers.
For example, Cruz et~al. (2007) identified 33 giants and 17 carbon stars among a
list of $\sim$200 ultracool dwarf candidates in the field.
However, they only used 2MASS $JHK_{\rm s}$ photometry for the selection.
The employment of at least one optical filter and astrometric and spectroscopic
information from the literature can help eliminate these contaminants in cluster
(and field) surveys.  

\item Photometric variability of red (sub-)giants in the background is an
additional source of contamination for surveys whose multiband images were taken
at different epochs (i.e., the colours of variable sources are not ``real'').
Many of such contaminants, although difficult to discard {\em a~apriori}, can be
identified by studying their light curves. 
Very often, the deep final image in a pass band of a survey is the combination
of a long series of exposures (this methodology, if accompanied with a suitable
dithering, allows to reach deeper magnitudes than in a single exposure of the
same total time).  
See Caballero et~al. (2004) and Scholz \& Eisl\"offel (2004) for examples of
photometric variability studies in a young open cluster based on such 
long~series. 

\item The major contributors to red contamination in deep surveys in
clusters are, therefore, field very low-mass stars and brown dwarfs with late
spectral types.  
The discovery of distant, late-type dwarfs in deep surveys is not
exceptional.
For example, Stanway et~al. (2008) found a population of M dwarfs at large
heliocentric distances in the GOODS fields.
Pirzkal et~al. (2005) also identified 18 M- and two L-type candidate
dwarfs in the {\em Hubble} Ultra Deep Field, that require further observations
for confirmation. 
Many more field L dwarfs have been found in the foregrounds of young open
clusters and associations (e.g., $\sigma$~Orionis: Caballero et~al. 2007; Upper
Scorpius: Lodieu et~al. 2008).
Cuby et~al. (1999) identified during an extragalactic survey the
most distant field T dwarf detected to date (\object{NTTDF~J1205--0744}), at an
heliocentric distance $d \sim$ 90\,pc, and with a magnitude $J \sim$ 20\,mag.
Kendall et~al. (2007) and Lodieu et~al. (2007a) recently estimated slightly
lower distances ($d \sim$ 50--80) for the intermediate T dwarfs
\object{ULAS~J145243.59+065542.9} and \object{ULAS~J223955.76+003252.6},
respectively.  
The actual status of \object{S\,Ori~70}, a $\sim$T6-type, $\sim$3\,$M_{\rm
Jup}$, planetary-mass object candidate towards Orion, is still under debate
(Zapatero Osorio et~al. 2002b, 2008; Mart\'{\i}n \& Zapatero Osorio 2003;
Burgasser et~al. 2004; Scholz \& Jayawardhana 2008). 
\end{itemize}

\subsection{The Galaxy thin-disc model}
\label{themodel}

Within the context of the ``standard'' Galactic model, the Galaxy can be
modelled by a double exponential, that represents the thin and the thick discs,
and a power law for the halo (e.g., a de Vacouleurs-type power law).
Classic works and reviews on Galactic structure were presented by, e.g., Bahcall
\& Soneira (1980), Gilmore \& Reid (1983), Majewski (1993) and Kroupa, Tout \&
Gilmore (1993).  
Recent observational works on Galaxy disc models from deep imaging surveys can
be found in, for example, Phleps et~al. (2005), Ryan et~al. (2005),
Karaali (2006), and Juri\'c et~al.~(2008). 

The volume densities of the axisymmetric thin and thick discs depend on the
galactocentric distance $R$ and on the height over or below the galactic plane,
$Z$.  
For an object of spectral type $i$, the expression for its spatial
density depending on heliocentric galactic coordinates ($l$, $b$) and
heliocentric distance ($d$) is usually written as follows: 

\begin{equation}
n_i(d,l,b) = n_{0,i} e^{-\frac{R(d,l,b)-R_\odot}{h_R}} e^{-\frac{|Z_\odot+d\sin{b}|}{h_Z}}
\label{eq.n_i}
\end{equation}

\noindent (e.g., Chen et~al. [2001] and references therein), where $n_{0,i}$ is
the star spatial density at the Galactic plane ($Z =$ 0) and at the solar
galactocentric distance ($R = R_\odot$), $Z_\odot$ is the height of the Sun over
the Galactic plane, and $h_R$ and $h_Z$ are the length (radial) and height
scales, respectively.
The exponential vertical distribution actually
comes from the approximation sech$^2 (|Z| / 2h_Z) \approx
e^{-|Z| / h_Z}$ (Gilmore, King \& van~der~Kruit 1990).
We discard using more complex forms for the vertical density profile (e.g.,
Zheng et~al. 2004), since additional parameters that characterize these forms
(e.g., $\beta$) are more poorly constrained than the exponential scale height
factors (see also the following discussion on the thin-to-thick disc
normalisation).
The galactocentric distance $R(d,l,b)$ of the object is, in its turn: 

\begin{equation}
R(d,l,b) = (R_\odot^2 + d^2 \cos^2{b} - 2 R_\odot d \cos{b} \cos{l})^{1/2}.
\label{eq.R}
\end{equation}

\noindent (e.g., Bahcall \& Soneira 1980 -- $R$ is measured along the Galactic
plane).
Assuming that the heliocentric distance, $d$, is much less than the solar
galactocentric distance, $d \ll R_\odot$ (see below for a justification), then
the galactocentric distance of the object can be approximated by:  

\begin{equation}
R(d,l,b) \approx R_\odot - d \cos{b} \cos{l}.
\label{eq.R.approx}
\end{equation}

\noindent The minus sign guarantees that $R < R_\odot$ when the line of sight
points to the Galactic Centre ($l$ = 0\,deg), and vice versa (we are using the
right-handed Cartesian galactocentic coordinate system in which the $X$ axis
points towards the Sun, the $Y$ axis in the anti-direction of Galactic rotation
and the $Z$ axis towards the North Galactic Pole).
In the linear approximation, Eq.~\ref{eq.n_i} can be rewritten as:  

\begin{equation}
n_i(d,l,b) \approx n_{0,i} e^{\frac{d\cos{b} \cos{l}}{h_R}} e^{-\frac{\pm Z_\odot \pm d\sin{b}}{h_Z}} 
= n_{0,i} e^{\frac{\mp Z_\odot}{h_z}} e^{-d (-\frac{\cos{b} \cos{l}}{h_R} \pm \frac{\sin{b}}{h_Z})},
\label{eq.n_i.approx}
\end{equation}

\noindent where the sign convention depends on whether $d\sin{b}$ is greater or
less than $-Z_\odot$ (i.e., whether the source is above or below the Galactic
plane). 
Defining two auxiliar variables that vary with the galactic coordinates,
$n_{A,i}$ and $d_B(l,b)$, Eq.~\ref{eq.n_i} results on a simple expression with
the heliocentric distance:  

\begin{equation}
n_i(d,l,b) \approx n_{A,i} e^{-\frac{d}{d_B(l,b)}},
\label{eq.n_i.AB}
\end{equation}

\noindent where the auxiliar variables are computed through:

\begin{equation}
n_{A,i} \equiv n_{0,i} e^{\mp \frac{Z_\odot}{h_Z}}
\label{eq.nA_i}
\end{equation}

\noindent and

\begin{equation}
\frac{1}{d_B(l,b)} \equiv -\frac{\cos{b} \cos{l}}{h_R} \pm \frac{\sin{b}}{h_Z}.
\label{eq.dB}
\end{equation}

   \begin{table}
      \caption[]{Main parameters of the Galaxy thin-disc model from Chen et~al.
      (2001).} 
         \label{parameters}
     $$ 
         \begin{tabular}{l c c c c}
            \hline
            \hline
            \noalign{\smallskip}
Parameter  	& $R_\odot$ 	& $Z_\odot$ 	& $h_R$ 		& $h_Z$ \\
	  	& (pc)		& (pc)		& (pc)			& (pc) \\
            \noalign{\smallskip}
            \hline
            \noalign{\smallskip}
Value    	& 8\,600$\pm$200& +27$\pm$4   	& 2\,250$\pm$1\,000	& 330$\pm$3 \\
            \noalign{\smallskip}
            \hline
         \end{tabular}
     $$ 
   \end{table}

In Table~\ref{parameters} we show the main parameters of the Galaxy thin-disc
model from Chen et~al. (2001).
These values suitably fit classic and recent determinations (e.g., Faber et~al.
1976; Bahcall \& Soneira 1984; Kerr \& Lynden-Bell 1986; Kuijken \& Gilmore
1989; Robin, Reyl\'e \& Cr\'ez\'e 2000; Pirzkal et~al. 2005; Juri\'c et~al. 2008
and references above).  
We will subsequently use only Chen et~al.'s values for consistency, even
though there are better estimates for some parameters (e.g., the solar
galactocentric distance -- Reid 1993; Eisenhauer et~al. 2003). 
From the constancy of the height scale of GKM-type dwarfs, as shown by Bahcall
\& Soneira (1980) and Gilmore \& Reid (1983), it seems reasonable to also use
the same height scale for M, L and T dwarfs.
Pirzkal et~al. (2005) measured $h_z$ = 400$\pm$100\,pc for late M and L dwarfs,
which is a little larger than ours, but still consistent.
While the vertical scale heights of early-type stars (O, B, A) are shorter, they
will not typically contribute to the photometrically selected sample and are
rare anyways.
 
The thick disc is more rarified and extended than the thin disc, with a
thin-to-thick disc normalisation factor in the solar vicinity of 13--6.5\,\%
(Chen et~al. 2001; Juri\'c et~al. 2008). 
This value interval is larger than previously reported, like the 2\,\% in
Gilmore (1984) or the lack evidence for a thick disc in Bahcall \& Soneira
(1984). 
According to theoretical cooling sequences (Burrows et~al. 1997; Chabrier et~al.
2000; Baraffe et~al. 2003), very old thick-disc substellar objects with very low
masses ($\mathcal{M} \ll$ 0.08\,M$_\odot$) may be dimmed to exceedingly faint
magnitudes, which would prevent their detection (thick-disc stars and brown
dwarfs are thought to be very old, with ages $\gtrsim$ 10\,Ga -- e.g., Fuhrmann
1998; Prochaska et~al. 2000; Feltzing, Bensby \& Lundstr\"om 2003). 
For example, an 0.05\,M$_\odot$-mass brown dwarf dims from $M_I$ = 17.2\,mag at
1\,Ga to $M_I$ = 20.5\,mag at 10\,Ga.
The same brown dwarf at that time intervals would have effective temperatures
typical of L5--7V and T7--9V, respectively.
Since very low-mass stars above the hydrogen burning limit (with late M and very
early L spectral types) keep a roughly constant luminosity in the main sequence
for tens gigayears, we expect that the ratio between L and T (and cooler) dwarfs
in the (relatively old) thick disc is less than in the (relatively young) thin
disc.
In other words, field late-M-, L- and T-type dwarfs of the solar neighbourhood
are younger, in average, than G-, K- and early M-type dwarfs (see Table~5 in
Zapatero Osorio et~al. 2007a).
There is a discussion in the negligibility of thick disk and halo brown dwarfs
in a given imaging sample in Pirzkal et~al. (2005).
Both the low thin-to-thick disc normalisation factor and the relative faintness
of thick-disc objects suggest that it is appropriate to use just the thin-disc
exponential for determining the density of late-type dwarfs in the Galaxy. 
The error in this assumption ($\lesssim$10\,\%) is smaller than, or of the
order of, the uncertainties in the determination of $R_\odot$, $Z_\odot$ and
$h_R$. 

\begin{figure}
\centering
\includegraphics[width=0.52\textwidth]{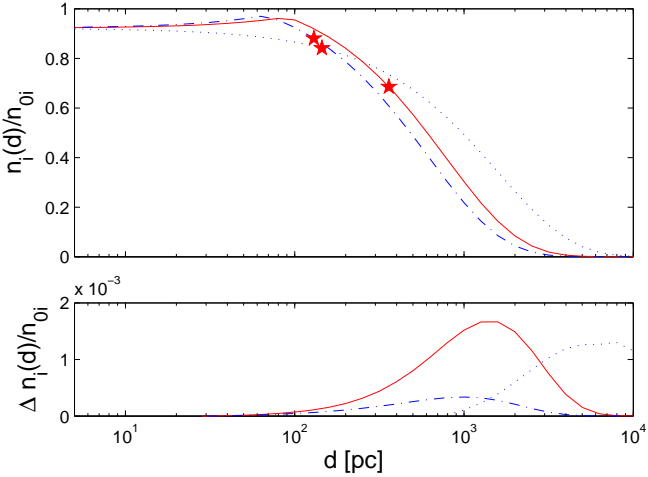}
\caption{{\em Top window:}
variation of the relative spatial density of stars and brown dwarfs of spectral
type~$i$ in the thin disc with the heliocentric distance from 5 to 10\,000\,pc
in the direction of three representative young clusters: Upper Scorpius
(dotted line), Pleiades (dash-dotted line), and $\sigma$~Orionis (--red-- solid
line). 
The location of the clusters are marked with a star `$\star$'~symbol.
{\em Bottom window:}
differences between the relative spatial densities using the correct
(Eq.~\ref{eq.R}) and approximate (Eq.~\ref{eq.R.approx})~expressions.}
\label{cumulos}
\end{figure}
%

Using the values in Table~\ref{parameters}, the ratio $n_{A,i} / n_{0,i} =
e^{\mp \frac{Z_\odot}{h_Z}}$ (Eq.~\ref{eq.nA_i}) equals 0.921$\pm$0.011 for $Z >
0$, and 1.085$\pm$0.013 for $Z < 0$ and is independent of the galactic
coordinates.  
However, the auxiliar variable $d_B(l,b)$ strongly depends on the galactic
longitude and latitude in the direction of observation and has a large interval
of~variation.

Fig.~\ref{cumulos} shows the suitability and accuracy of the linear
approximation of $R(d,l,b)$. 
The differences between the relative spatial densities of stars and brown dwarfs
of spectral type~$i$ in the thin disc, $n(d)/n_0$, using the correct
(Eq.~\ref{eq.R}) and approximate (Eq.~\ref{eq.R.approx}) expressions are not
larger than 0.2\,\% in the direction of three representative open clusters,
whose galactic coordinates and distances are given in Table~\ref{clusters}. 
The differences are two orders of magnitude lower ($<$ 0.002\,\%) at
heliocentric distances $d <$ 100\,pc, where the expected number of late-type
dwarf contaminants is maximum. 
These differences are very small in comparison with the errors in the
determination of the main parameters of the Galaxy disc model and, as we will
see later, the spatial densities for each spectral type.
Besides, the linear $R(d,l,b)$ approximation enormously simplifies next
analysis.

All $n(d)/n_0$ curves in the top panel of Fig.~\ref{cumulos} show the same local
density at null heliocentric distance, which is for an object of spectral
type~$i$:  

\begin{equation}
n_i(R = R_\odot,Z = Z_\odot) = n_{0,i} e^{-\frac{Z_\odot}{h_z}} \equiv n_{\odot,i}.
\label{eq.nodot}
\end{equation}

\noindent Upper Scorpius is in the Galactic north hemisphere, like the Sun, so
the line of sight of the search does not cross the Galactic plane.
Thus, in contrast to the Pleiades and $\sigma$~Orionis, the Upper Scorpius
relative star density has not a maximum at $Z =0$ (and $n_{A,i}(d,b)$ and
$d_{B,i}(d,l,b)$ do not change sign). 
For the Pleiades and $\sigma$~Orionis, the peak of $n(d)/n_0$ at $d\sin{b} =
-Z_\odot$ (when the line of sight intersects the Galactic plane) is not exactly
1.00, but there is an error of about 3\,\% because the galactic coordinate
system ``is necessarily based on a galactic plane through the Sun'', which is
the H~{\sc i} principal plane (Blaauw et~al. 1960). 
Finally, the spatial density in the direction to Upper Scorpius is smaller than
in the direction to the other clusters at relatively small heliocentric
distances, but it gets larger at more than about 250\,pc because the OB
association is roughly in the direction of the Galactic Centre. 

   \begin{table}
      \caption[]{Heliocentric distances and galactic coordinates of three
      representative young open clusters.} 
         \label{clusters}
     $$ 
         \begin{tabular}{l c c c}
            \hline
            \hline
            \noalign{\smallskip}
Region  		& $d$	& $l$ 	& $b$ 		\\
			& (pc)	& (deg) & (deg) 	\\
            \noalign{\smallskip}
            \hline
            \noalign{\smallskip}
Upper Scorpius    	&  145  & 350 	&   +20 	\\
$\sigma$ Orionis    	&  385  & 207 	&  --17 	\\ 
Pleiades   		&  130  & 167 	&  --24 	\\ 
            \noalign{\smallskip}
            \hline
         \end{tabular}
     $$ 
   \end{table}

\subsection{Integrated number of contaminants}
\label{numberofcontaminants}

\subsubsection{The pseudo-algorithm}
\label{thepseudoalgorithm}

\begin{figure}
\centering
\includegraphics[width=0.52\textwidth]{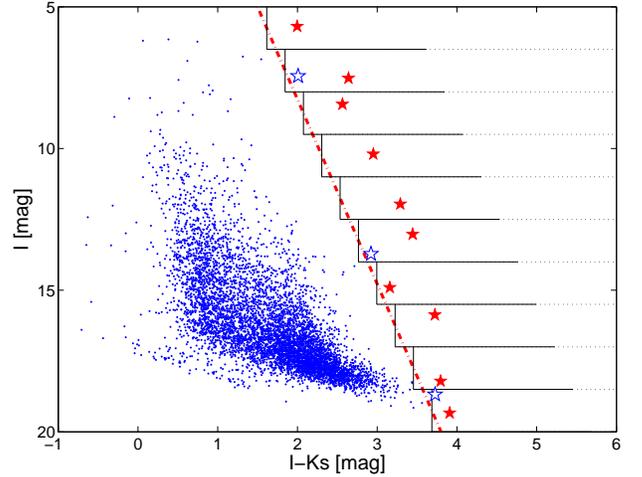}
\caption{Simulated $I_{\rm SSA}$ vs. $I_{\rm SSA}-K_{\rm s, 2MASS}$
colour-magnitude diagram of a hypothetical cluster survey.
{\em (Red) filled stars (10):} hypothetical confirmed cluster members with
spectroscopic signatures of youth;
{\em (blue) open stars (3):} hypothetical cluster member candidates based on
photometry;
{\em (blue) small dots:} probable fore- and background sources;
{\em (red) thick dot-dashed line:} selection criterion used for separating
cluster members and candidates from probable fore- and background sources;
{\em (black) thin solid/dotted lines:} ``strips'' where the number of
contaminants are computed.
The height of the strips were set to 1.5\,mag for clarity.
Small dots and open stars are actual data from a multiband ($BRI_{\rm
SSA}JHK_{\rm s, 2MASS}$) 90\,arcmin-radius survey centred on \object{HD~221356}
(aka K\"o\,3\,A), an F8V star with an M--L-type binary companion in a very wide
orbit (Gizis et~al. 2000; Caballero 2007).
The mean sequence of the hypothetical cluster is parallel to the thick
dot-dashed line, shifted 0.5\,mag to the red.
Filled stars were randomly generated by adding a 0.5\,mag-amplitude noise to the
mean cluster sequence (i.e., the selection criterion is $\sim$1\,$\sigma$
shifted to the blue with respect to the mean cluster sequence).
[SSA: SuperCOSMOS Science Archive (Hambly et~al. 2001); 2MASS: Two-Micron All
Sky Survey (Skrutskie et~al.~2006)]. 
} 
\label{simulatedc-mdiagram}
\end{figure}

As introduced in Section~\ref{introduction}, the number of possible field-dwarf
contaminants in a survey can be computed from the corresponding colour-magnitude
diagram.
In Fig.~\ref{simulatedc-mdiagram}, we illustrate the computation of contaminants
based on a {\em simulated} colour-magnitude diagram.
The pseudo-algorithm of integration is identical for any combination of ``red''
(e.g.,~$K_{\rm s}$) and ``blue'' (e.g.,~$I$) passbands.

The general procedure to compute the integrated number of field-dwarf
interlopers consists~on:  

\begin{itemize}
\item[1.] {\em Defining horizontal ``strips'' in the colour-magnitude diagram.}
They are discrete rectangular regions in the diagram that broke the area of
selected sources in pieces for a suitable integration. 
The blue border of each strip is defined by the selection
criterion; 
the choice of the red border is, however, rather subjective, because observers
tend to select as cluster member candidates {\em all} the sources to the red of
the selection criterion. 
A reasonable choice of the red border is the colour of the reddest source in
each magnitude interval, but it can be extended up to very red colours
(indicated with dotted lines in Fig.~\ref{simulatedc-mdiagram}).
The height of the strips (i.e., the size of the integration step) depends on the
required accuracy; 
a width of $\sim$0.5\,mag can be a compromise between a fast computation and
fidelity in the reproduction of the shape of the selection criterion (in this
case, a straight line). 
This height cannot be less than the typical photometric uncertainty.

\item[2.] {\em Determining the spectral types of the possible interlopers in
each strip.}
They are derived from the colours at the strip borders and from a spectral
type-colour relation.
See below for caveats on this determination.

\item[3a.] {\em Setting the minimum and maximum heliocentric distances $d_1$ and
$d_2$} at which an interloper of a given spectral type would lie within each
strip, based on the corresponding absolute magnitudes, $M_\lambda$.  
The computation of $M_\lambda$ must account for the apparent magnitude
$m_\lambda$, the heliocentric distance $d$ and the interstellar extinction
$A_\lambda$ (the error in the $M_\lambda$-spectral type relation is
considered in item~\#2).  

\item[3b.] {\em Setting the heliocentric distance $d_*$ at which the line of
sight crosses the Galactic plane} in case the minimum and maximum heliocentric
distances are in different Galactic hemispheres. 

\item[4.] {\em Integrating the spatial density} between the minimum and maximum
heliocentric distances to get the number of interlopers of a given spectral type
per surface unit in each~strip. 

\item[5.] {\em Summing the number of interlopers} of all the considered spectral
types and in all the strips.

\item[6.] {\em Multiplying the value by the survey area},~$\mathcal{S}$, to get
the integrated number of contaminants.
\end{itemize}

Regarding item~\#2, a field dwarf with intrinsic colour bluer than the
selection limit can still lie in the selected region of the colour-magnitude
diagram because of photometric error. 
Moreover, a spectral type does not correspond to one absolute magnitude and one
colour value but to a range of magnitudes and colours. 
It is necessary to take these effects into account when calculating the
contamination, otherwise the result will be underestimated.
One  way for accounting for the photometric scattering of the data (including
the broadening of the main sequence and the survey photometric uncertainties)
may be choosing one earlier spectral subtype of the bluer interloper than
expected from the selection criterion. 
At the late considered spectral types, a variation of one subtype corresponds to
an average colour variation of, e.g., $\Delta (I-J) \approx$ 0.2--0.3\,mag (see
below), which may be a too large photometric uncertainty for a typical survey.
An alternative solution is simply to shift bluewards the selection criterion
actually used in the colour-magnitude diagram by certain amount
($\sim$0.05--0.10\,mag).  
This amount would depend on the survey, magnitude range, and considered spectral
types.

Once the main parameters (Table~\ref{parameters}) and the exponential form of
the Galaxy thin-disc model (Eq.~\ref{eq.n_i}) are known, as well as the spatial
densities, colours and absolute magnitudes of the possible contaminants as a
function of spectral type (Section~\ref{thecontaminants}), the integration can
be carried~out.

\subsubsection{The integration}
\label{theintegration}

In this Section we provide the necessary tools to integrate the exponential
spatial density in the linear approximation (Eq.~\ref{eq.n_i.AB}). 
In a typical photometric search program, assuming that such a search does not
span an excessively wide area (i.e., smaller than a few tens of square degrees),
the search volume can be approximated as a cone with an apex length equal to the
heliocentric distance of a given source detectable within the photometric depth,
and a base corresponding to the projected area on the sky of the search,
$\mathcal{S}$.  
The axis of the cone is defined by the line of sight (i.e., the galactic
latitude and longitude of the centre of survey area). 
The use of a cone makes the computation of late-type contaminants much easier,
since it is not necessary to know the actual shape of the~survey.

The number of late-type dwarfs of spectral type $i$ that contaminate the survey
comes from the integration in a truncated cone of volume $V$ with a non-uniform
density $n(\mathbf{r})$:  

\begin{eqnarray}
 & N_i = \int n_i(\mathrm{d}) {\rm d}V = \int_{z} \int_{\phi} \int_{\alpha}
 n_i(z,\phi,\alpha) {\rm d}V,  
\end{eqnarray}

\noindent where the volume differential in the coordinate system of the cone
is $\frac{z^2 \sin{\alpha}}{\cos^3{\alpha}} {\rm d} z {\rm d} \phi {\rm d}
\alpha$ (being $z$ the regular z-coordinate, $\phi$ the angle of a
projected vector with the positive X-axis, measured counterclockwise, and
$\alpha$ the angle between the vertical axis, the cone vertex and any other
point of the cone). 
When the galactic latitude and longitude of the field of view are known, the
density $n(\mathbf{r})$ depends only on the heliocentric distance, $d$
(the ``height'' of the cone).
Using the simplified expression of the density of thin-disc dwarfs
(Eq.~\ref{eq.n_i.AB}), the integration between the heliocentric distances $d_1$
and $d_2$ provides: 

\begin{eqnarray}
 & N_i = \int_{d_1}^{d_2} n_i(z) z^2 {\rm d} z \int_{0}^{2\pi} {\rm d}
  	\phi \int_{0}^{\arctan{(b/h)}} \frac{\sin{\alpha}}{\cos^3{\alpha}}
	{\rm d} \alpha = \nonumber \\
 & = \pi \left( \frac{b}{h} \right)^2 \int_{d_1}^{d_2} n_{A,i}
	 e^{-\frac{z}{d_B}} z^2 {\rm d} z,   
\end{eqnarray}

\noindent In the pyramid-cone approximation of height $h$ and base radius $b$,
the value $\pi \left(\frac{b}{h} \right)^2$ is simply the angular area of the
search, $\mathcal{S}$, which must be expressed in rad$^2$. 
If $d_1$ and $d_2$ are each in a Galactic hemisphere (i.e., the line of
sight crosses the Galactic plane), then the integral is detached into two~parts:

\begin{eqnarray}
 & N_i = \mathcal{S} \left( n_{A_{+,i}} \int_{d_1}^{d_*}
	e^{-\frac{z}{d_{B_+}}} z^2 {\rm d} z + n_{A_{-,i}} \int_{d_*}^{d_2}
	e^{-\frac{z}{d_{B_-}}} z^2 {\rm d} z \right),   
\end{eqnarray}

\noindent where $d_*$ is the heliocentric distance at which $n(d)$ has a maximum
($Z$ = 0) and the $n_A$ and $d_B$ auxiliar variables change behaviour.
The exponential integral has a primitive, which is:

\begin{eqnarray}
 & \int_{d_i}^{d_j} e^{-\frac{z}{d_B}} z^2 {\rm d} z = 
 	\left[ (-1) e^{-\frac{z}{d_B}} (d_B z^2 +2 d_B^2 z + 2 d_B^3)
	\right]_{d_i}^{d_j} \equiv \\
 & \equiv I(d_B; d_i, d_j).
\end{eqnarray}

\noindent Hence, the final expression for the total number of dwarfs in a truncated cone
with the minor and major bases at heliocentric distances $d_1$ and $d_2$,
respectively, is:

\begin{eqnarray}
 & N_i(r_1,r_2) = \mathcal{S} \left[ n_{A_{+,i}} I(d_{B_+}; d_1, d_*) +
 n_{A_{-,i}} I(d_{B_-}; d_*, d_2) \right].  
\label{thetotalnumber}
\end{eqnarray}

\subsection{Possible late-type contaminants}
\label{thecontaminants}

\begin{figure}
\centering
\includegraphics[width=0.49\textwidth]{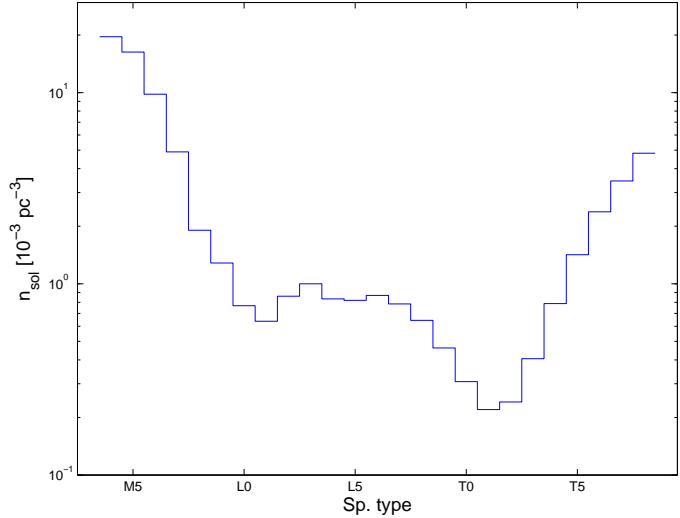}
\caption{Local spatial density of ultracool dwarfs as a function of spectral
type from the data in Table~\ref{MLT}.}
\label{densityofUCDs}
\end{figure}

We have studied the contribution to contamination in photometric searches 
for very red sources by field dwarfs with spectral types from M3V to T8V. 
This election is based, on the one hand, on the earliest spectral type of a
brown dwarf in a $\sim$1\,Ma-old star-forming region, which is about M5, and the
large-amplitude photometric variability that can be present in the brightest
young substellar objects, which can be as large as 0.7\,mag (Caballero et~al.
2006). 
The maximum colour variation observed could be even larger than the amplitude of
photometric variability.
Because of that, if the surveys in each passband are not simultaneous, very
young variable brown dwarfs can display colours of objects a few sub-types
earlier than its actual spectral type. 
On the other hand, the late-type bound is set by the latest-type brown dwarfs
observed to date (e.g., Burgasser et~al. 2000; Tinney et~al. 2005; Looper,
Kirkpatrick \& Burgasser 2007; Lodieu et~al. 2007a; Warren et~al. 2007).  
The total number of contaminants in the survey will be the sum of the number of
field dwarfs of spectral type $i$ in the truncated cones of~search:

   \begin{table}
      \caption[]{Characteristics of possible late-type field dwarf
      contaminants from the literature$^{a}$.} 
         \label{MLT}
     $$ 
         \begin{tabular}{l c c c c c}
            \hline
            \hline
            \noalign{\smallskip}
Spectral	& $M_I$		& $I-J$		& $J-K_{\rm s}$ 	& $n_\odot$		& $d_{I=25}^{b}$ 	\\  
type   		& (mag)  	& (mag)		& (mag)			& (10$^{-3}$ pc$^{-3}$)	& (pc)	   	\\ 
            \noalign{\smallskip}
            \hline
            \noalign{\smallskip}
M3--4\,V	&  9.07  	& 1.83	& 0.84	&19.6		& 7700 \\ %
M4--5\,V 	& 10.14 	& 1.80	& 0.86	&16.3	       	& 5700 \\ %
M5--6\,V  	& 11.91  	& 2.47	& 0.90	& 9.81         	& 3100 \\ %
M6--7\,V  	& 12.90  	& 2.72	& 0.93	& 4.90         	& 2200 \\ %
M7--8\,V  	& 14.16  	& 3.24	& 0.98	& 1.907        	& 1300 \\ %
M8--9\,V  	& 14.68  	& 3.54	& 1.06	& 1.285        	& 1100 \\ %
M9--L0\,V 	& 15.23  	& 3.80	& 1.12	& 0.768        	&  840 \\ %
L0--1\,V  	& 15.27  	& 3.55	& 1.27	& 0.637        	&  820 \\ %
L1--2\,V  	& 15.69  	& 3.50	& 1.38	& 0.861        	&  680 \\ %
L2--3\,V  	& 16.19  	& 3.62	& 1.43	& 1.000	       	&  550 \\ %
L3--4\,V  	& 16.67  	& 3.72	& 1.49	& 0.834        	&  450 \\ %
L4--5\,V  	& 17.14  	& 3.81	& 1.54	& 0.819        	&  360 \\ %
L5--6\,V  	& 17.61  	& 3.90	& 1.60	& 0.869        	&  290 \\ %
L6--7\,V  	& 18.07  	& 4.06	& 1.66	& 0.785        	&  240 \\ %
L7--8\,V  	& 18.52  	& 4.12	& 1.71	& 0.644        	&  190 \\ %
L8--9\,V  	& 18.90  	& 4.22	& 1.81	& 0.462        	&  160 \\ %
L9--T0\,V 	& 18.95  	& 4.40	& 1.73	& 0.308        	&  160 \\ %
T0--1\,V  	& 19.01  	& 4.63	& 1.58	& 0.220        	&  160 \\ %
T1--2\,V  	& 19.04  	& 4.94	& 1.17	& 0.241        	&  150 \\ %
T2--3\,V  	& 19.04  	& 5.19	& 0.81	& 0.406        	&  150 \\ %
T3--4\,V  	& 19.07  	& 5.23	& 0.51	& 0.788        	&  150 \\ %
T4--5\,V  	& 19.22  	& 5.01	& 0.29	& 1.42 	       	&  140 \\ %
T5--6\,V  	& 19.61  	& 4.76	& 0.11	& 2.38         	&  120 \\ %
T6--7\,V  	& 20.35  	& 5.25	& 0.00	& 3.45         	&   85 \\ %
T7--8\,V  	& 21.60  	& 5.50	&--0.06	& 4.82         	&   48 \\ %
            \noalign{\smallskip}
            \hline
         \end{tabular}
     $$ 
\begin{list}{}{}
\item[$^{a}$] 
The sources of the $M_I$ absolute magnitude, $I-J$ and $J-K_{\rm s}$ colours,
and spatial densities are obtained or derived by us from data in the literature
(Kirkpatrick et~al. 1994; Dahn et~al. 2002; Hawley et~al. 2002; Cruz et~al.
2003; Tinney et~al. 2003; Vrba et~al. 2004; Bochanski et~al. 2007; Burgasser
2007; West et~al. 2008). 
See the text for further details.
\item[$^{b}$] 
$d_{I=25}$ is the heliocentric distance of a dwarf with an apparent magnitude
$I$ = 25.0\,mag, assuming an interstellar extinction proportional to the
heliocentric distance. 
At the end of Section~\ref{thecontaminants} there is a brief discussion on this
assumption.
\end{list}
   \end{table}
%
%

\begin{equation}
N = \sum_{i=1}^{\rm Num.} N_i(r_{1,i},r_{2,i}), 
\label{eqn.N}   
\end{equation}

\noindent ($i$ =M3--4V, M4--5V, M5--6V... T7--8V).

The Johnson $I$-band absolute magnitudes, $I-J$ and $J-K_{\rm s}$ colours ($J$
and $K_{\rm s}$ in the 2MASS system) and {\em local} spatial densities from the
literature of the 25 considered intervals of spectral types are given in
Table~\ref{MLT}. 
For the M3--L1 dwarfs, we used the tabulated values from West et~al. (2008)
of $i_{\rm SDSS}-J$, then calculated the $I - i_{\rm SDSS}$ corrections using
the template spectra of Bochanski et~al. (2007) to get the $I-J$ colours. 
Absolute $I$-band magnitudes were computed from absolute $J$-band magnitudes
compiled by Hawley et~al. (2002) and using our $I-J$ values. 
$J-K_{\rm s}$ colours are also from West et~al. (2008).
For the L8--T0 dwarfs, $J-K_{\rm s}$ values are from {\tt DwarfArchives.org},
taking average of objects with uncertainties $\delta(J-K_{\rm s}) <$ 0.1\,mag
and not known to be binary.  
The absolute $I$ and $I-J$ colours are based on the polynomial relations
of Tinney, Burgasser \& Kirkpatrick (2003 -- note the backtrack in $I-J$ colours). 
The remaining $IJK_{\rm s}$ photometry has been taken from Dahn et~al. (2002)
and Vrba et~al. (2004). 

The local spatial densities ($n_\odot$) and the spatial densities at $R=R_\odot$
and $Z=0$ ($n_0$) are related through Eq.~\ref{eq.nodot}.
A pictorical representation of $n_\odot$ versus spectral type is given in
Fig.~\ref{densityofUCDs}. 
Spatial densities of M3--7 and M7--L0 dwarfs are measured values from
Kirkpatrick et~al. (1994) and Cruz et~al. (2003), respectively.  
For L0--T8 dwarfs, we used predicted values from Burgasser (2007) based on a
Monte Carlo simulation assuming a mass function dN/dM $\propto$ M$^{-0.5}$
(similar to estimates from various star forming regions -- e.g., Luhman et~al.
2000; Caballero et~al. 2007), a normalisation of 0.0037 pc$^{-3}$ over the range
0.09--0.10\,M$_\odot$ (Reid et~al. 1997), a binary fraction of 10\,\% and a
binary mass ratio distribution of $P(q) \propto q^4$ (where $q \equiv$
M$_2$/M$_1$). 
These values are consistent with empirical space density estimates by Burgasser
et~al. (2002), Cruz et~al. (2007) and Metchev et~al. (2008), and predicted
densities by Deacon \& Hambly (2006), although they differ significantly from
predictions by Nakajima (2005) for L5--T7 dwarfs.

Using the expression $I-M_I = 5 \log{d}-5+A_I$, a deep survey with a limiting
magnitude $I_{\rm lim}$ = 25.0\,mag would be able to detect a T5--6 
dwarf at an heliocentric distance $d \sim$ 120\,pc and an M5--6 dwarf at
about 3.1\,kpc to the Sun (see last column in Table~\ref{MLT}). 
In the direction of the Pleiades, such an hypothetical M-type ``field'' dwarf
would be located at about 1.9\,kpc below the Galactic plane at almost $z = 6
\times h_z$. 
In this location, the spatial density of thin-disc objects of any spectral type
is about 10$^{-3}$ times the local density $n_\odot$. 
It is evident from this example the necessity of the integration along the cone
of search using a non-uniform density.
Following this assumption, we should also use an interstellar extinction that
does {\em not} linearly increase with heliocentric distance.
Instead, it should be proportional to the actual interstellar dust content. 
In our model, we assume that the Galactic dust content has much wider length and
height scales than for~stars.

\section{A practical application: 
L and T dwarf contamination in $\sigma$~Orionis} 

\subsection{Preliminaries} 

The $\sigma$~Orionis cluster ($\sim$3\,Ma, $\sim$385\,pc) contains one of the
best known substellar populations, from the hydrogen burning mass limit down to
a few Jupiter masses (see a bibliographic review in Caballero 2008b).
It could be the most favourable site known to date for studying the opacity mass
limit for formation of objects via fragmentation in molecular clouds, that may
lie just below 0.003\,$M_\odot$ (Rees 1976; Bate, Bonnell \& Bromm 2003 and
references therein).
Even accounting for the L-type cluster members and candidates in the Pleiades
(Mart\'{\i}n et~al. 1998; Dobbie et~al. 2002b; Moraux et~al. 2003; Bihain et~al.
2006), Chamaeleon and \object{Lupus} (Allers et~al. 2006;
Jayawardhana \& Ivanov 2006), $\sigma$~Orionis contains by far the largest
number of L-type objects {\em with} spectroscopy (12 -- Zapatero Osorio et~al.
1999, 2000; Mart\'{\i}n et~al. 2001; Barrado y Navascu\'es et~al. 2001, 2003).
While there is reasonable criticism on the cluster membership of some of the
faintest objects (e.g., \object{S\,Ori~47}, L1.5$\pm$0.5 -- McGovern et~al.
2004), severeal of them are known to be extremely young based on near-infrared
excess and extraordinary H$\alpha$ emission of up to
--700\,\AA~(\object{S\,Ori~55}, \object{S\,Ori~71} -- Barrado y Navascu\'es
et~al. 2002; Zapatero Osorio et~al. 2002a; Caballero et~al. 2007) or flux excess
in the IRAC {\em Spitzer} bands (Zapatero Osorio et~al. 2007b; Scholz \&
Jayawardhana 2007).
Besides, $\sigma$~Orionis also possesses the only extremely young T-type
cluster member candidate identified so far (S\,Ori~70; it was introduced in
Section~\ref{introduction} -- Casewell et~al. 2007 and Bouvier et~al. 2008
have recently identified T dwarf candidates and cluster members in the Pleiades
[$\sim$120\,Ma] and the Hyades [$\sim$600\,Ma]). 

The contamination by field late-M dwarfs in the cluster is well bounded.
Caballero et~al. (2007) identified 30 objects as bona fide $\sigma$~Orionis
members with features of extreme youth amongst 49 candidate cluster members
fainter than $I$ = 16\,mag selected from an $I$ vs. $I-J$ diagram.  
They also reported the detection of two L-type field dwarf candidates in the
direction of the cluster, even fainter (and  much more distant) than Cuby
et~al. (1999)'s T~dwarf. 
The ratio between very low-mass confirmed and candidate cluster members in
$\sigma$~Orionis is comparable to that in the Orion Nebula Cluster (Lucas et~al.
2001; Slesnick, Hillenbrand \& Carpenter, 2004), but much larger than in other
older open clusters like the Praesepe (Pinfield et~al. 1997, 2003; Adams et~al.
2002; Gonz\'alez-Garc\'{\i}a et~al. 2006) or the Hyades (Reid \& Hawley 1999;
Gizis, Reid \& Monet 1999; Dobbie et~al. 2002a).
In any case, some of the faintest substellar member candidates in
$\sigma$~Orionis have no astrometric confirmation or irrefutable youth
signatures. 
Therefore, while there are no new data on cluster membership, it is
necessary to accurately determine the number of possible fore- and background
contaminants among selected photometric cluster member candidates for
investigating the cluster Initial Mass Function.  
We have applied our Galaxy and late-type dwarf data and mathematical tools to
estimate the number of fore- and background L and T dwarfs as a function
of depth in a hypothetical survey area of 1\,deg$^2$ ($\mathcal{S} \approx$
0.000\,30\,rad$^2$) in the $\sigma$~Orionis cluster. 
 
We have assumed that the cluster member selection is exclusively based on $I-X$
colours (Johnson $I$), being $X$ a near-infrared passband (e.g., $JHK_{\rm s}$).
The $X$-band observations are deep enough to detect within the completeness all
the L- or T-type object identified in the $I$-band images\footnote{For example,
the completeness magnitudes in the representative survey by Caballero et~al.
(2007) were $I_{\rm compl.}$ = 23.3\,mag, $J_{\rm compl.}$ = 20.6\,mag, allowing
the detection of all the objects within the completeness in the optical redder
than $I-J$ = 2.7\,mag ($\sim$M7V).}. 
The combination of a red optical filter and a near-infrared one is an
effective way (i.e., in telescope time, in simplicity) for selecting very faint
red objects in young open clusters.
Although an $I-K_{\rm s}$-based search would be better for identifying
late-L-type objects, the increase in thermal background at 2\,$\mu$m of
near-infrared detectors favours the use of the $J$ passband, especially for
T-type objects (that have quite blue $J-K_{\rm s}$ colours);
a compromise might be found at an $I-H$ survey.
Likewise, imaging at passbands bluewards of $I$ is prohibitive due to the
extremely dimness of ultracool cluster member candidates at such wavelenghts
(e.g., a cluster member with $I$ = 24\,mag is expected to have $R \approx$
27--28\,mag). 
Finally, the number of interlopers with pure optical ($R-I$, $I-Z$) and pure
infrared ($J-K_{\rm s}$) passbands is larger than with a combination of red
optical and near-infrared filters, because of the larger slope of the spectral
energy distribution of young late-type objects at the optical/infrared boundary
(i.e., redder $I-J$, $I-K_{\rm s}$ colours).

As explained in Section~\ref{thepseudoalgorithm}, one of the steps of the
integration of the number of contaminants is setting the minimum and maximum
heliocentric distances $d_1$ and $d_2$ specified in Eq.~\ref{thetotalnumber}.
For the computation of the heliocentric distance of an object of apparent and
absolute magnitudes $I$ and $M_I$, one must find the root of the smooth function
$f(d) = 5 \log{d}-5+A_I-(I-M_I)$, where the extinction $A_I = 0.482 A_V$ is
proportional to the distance ($A_I = a d$) and $A_V = 3.09 E(B-V)$
(Rieke \& Lebofski 1985).
The constant $a$ depends on the colour excess $E(B-V)_\star$ and distance
$d_\star$ towards $\sigma$~Orionis (e.g., Lee 1968; Caballero 2008a).
Quantitatively: $E(B-V) = E(B-V)_\star (d / d_\star)$.
To find the root, we have used the Newton method ($d_{n+1} = d_n - f(n)/f'(n)$,
$n$ = 0, 1, 2...), that quickly converges in 3--6 iterations with a suitable
initial value of~$d_0$.

   \begin{table}
      \caption[]{Number of expected field L- and T-type dwarfs in a 
      1\,deg$^2$-area survey towards $\sigma$~Orionis as a function of 
      the Johnson $I$-band magnitude and spectral type interval$^{a}$.}
         \label{contamination}
     $$ 
         \begin{tabular}{l c c c c}
            \hline
            \hline
            \noalign{\smallskip}
$\Delta I$ 	& L0--5V			& L5--T0V			& T0--5V			& T5--8V \\
(mag)		& 				& 				& 				&	\\
            \noalign{\smallskip}
            \hline
            \noalign{\smallskip}
21.0--22.0	& 1.1$^{+0.5}_{-0.4}$		& 0.049$^{+0.034}_{-0.019}$	& 0.012$^{+0.009}_{-0.005}$	& 0.008$^{+0.007}_{-0.003}$	\\ 
22.0--23.0	& 3.9$^{+1.8}_{-1.4}$		& 0.19$^{+0.14}_{-0.08}$	& 0.05$^{+0.03}_{-0.02}$	& 0.030$^{+0.027}_{-0.014}$	\\ 
23.0--24.0	& 12$^{+5}_{-4}$		& 0.7$^{+0.5}_{-0.3}$		& 0.18$^{+0.14}_{-0.08}$	& 0.12$^{+0.11}_{-0.06}$	\\ 
24.0--25.0	& 36$^{+14}_{-13}$		& 2.5$^{+1.7}_{-1.0}$  		& 0.7$^{+0.5}_{-0.3}$		& 0.5$^{+0.4}_{-0.2}$		\\ 
25.0--26.0	& 90$^{+30}_{-30}$		& 8$^{+4}_{-3}$			& 2.5$^{+1.8}_{-1.1}$		& 1.8$^{+1.5}_{-0.8}$		\\ 
26.0--27.0	& 180$^{+60}_{-80}$		& 24$^{+14}_{-10}$		& 8$^{+5}_{-3}$			& 6$^{+6}_{-3}$			\\ 
27.0--28.0	& 270$^{+100}_{-130}$		& 60$^{+30}_{-30}$		& 25$^{+16}_{-11}$		& 20$^{+17}_{-10}$		\\ 
28.0--29.0	& 300$^{+140}_{-170}$		& 130$^{+60}_{-60}$		& 60$^{+40}_{-30}$		& 60$^{+50}_{-30}$		\\ 
            \noalign{\smallskip}
            \hline
         \end{tabular}
     $$ 
\begin{list}{}{}
\item[$^{a}$] 
Poissonian localized overdensities are not taken into~account.
\end{list}
   \end{table}

In Table~\ref{contamination} we summarize our results and provide the number of
L0--T8V-type dwarfs in 1.0\,mag-width ``strips''. 
As discussed below, not all the late-type field dwarfs in the survey area
are true contaminants (e.g., L-type dwarfs do not contaminate the
$\sigma$~Orionis $I$ vs. $I-J$ colour-magnitude diagram at $I \gtrsim$
24.5\,mag). 
The uncertainties come from the errors in the parameters of the thin disc
(Table~\ref{parameters}), the cluster heliocentric distance ($d$ =
330--440\,pc), the spectral type-absolute magnitude relation, the colour excess,
and the local spatial densities.  
We use conservative errors of 0.2\,mag and 0.02\,mag for the spectral
type-$M_I$ relation and the interstellar extinction, respectively.
Variations in the solar galactocentric distance $R_\odot$ barely affect the
results.
The most important contributors of error are the radial scale length $h_r$, that
is the Galactic parameter with the largest intrinsic uncertainty ($>$~40\,\%),
and the local spatial densities.
The mass function index that Burgasser (2007) used
for deriving $n_\odot$ ($\alpha$ = +0.5; $dN/dM \propto M^{-\alpha}$) may
be too large according to recent observational results (Lodieu et~al. 2007b;
Metchev et~al. 2008).  
A flatter substellar mass function ($\alpha \approx$ 0.0) would lead to
lower local spatial densities and, therefore, lower contamination rates.
However, the choice for Burgasser (2007)'s single power-law mass function (from
where L0--T8V local densities in Table~\ref{parameters} were obtained) seems
more reasonable to us, since it explains the raise of the mass function found by
many other authors in the planetary regime (see Section~\ref{introduction}).
Besides, Metchev et~al. (2008)'s values are consistent with Burgasser
(2007)'s ones within the 95\,\% confidence limits.
Maximum differences with other determinations of substellar densities in the
literature are of the order of 20\,\%.
Our estimation of the uncertainties in $n_\odot$ is, however, more
conservative: maximum and minimum surface densities come from using very
different indices of the local mass function, $\alpha$ = 0 and +1 (see
table~5 in Burgasser 2007).
The error in $n_\odot$ strongly depends on the spectral type, and ranges from
$\sim$12\,\% at L0--1V to $\sim$50\,\% at T7--8V. 

Our computations surpass and complement the estimate of contamination in the
$\sigma$~Orionis by Caballero et~al.~(2007). 
The number of interlopers shown in Table~\ref{contamination} are consistent
with those given in Caballero et~al. (2007) accounting for the different survey
area, input absolute magnitudes $M_I$ and local densities $n_\odot$ (especially
in the L5--T7V interval), and width of strips (i.e., $I-J$ colour).  
They estimated that $\sim$4 L-type dwarf contaminants ($\sim$2 early,
$\sim$2 late) populated their least-massive bin of the Initial Mass Function,
that contained 11 planetary-mass object candidates (i.e., the contamination rate
is 36\,\%); our results support those calculations.  
A coarse extrapolation towards fainter magnitudes of the surface densities
provided by Burgasser (2007) also matches our results (he did not account,
however, for the Galactic structure).
Similar values as in Table~\ref{contamination} are expected for 1\,deg$^2$-wide
surveys of any kind (young open clusters, extragalactic) at intermediate
galactic and high latitudes of the same magnitude~depth.

\subsection{Discussion} 

\begin{figure}
\centering
\includegraphics[width=0.49\textwidth]{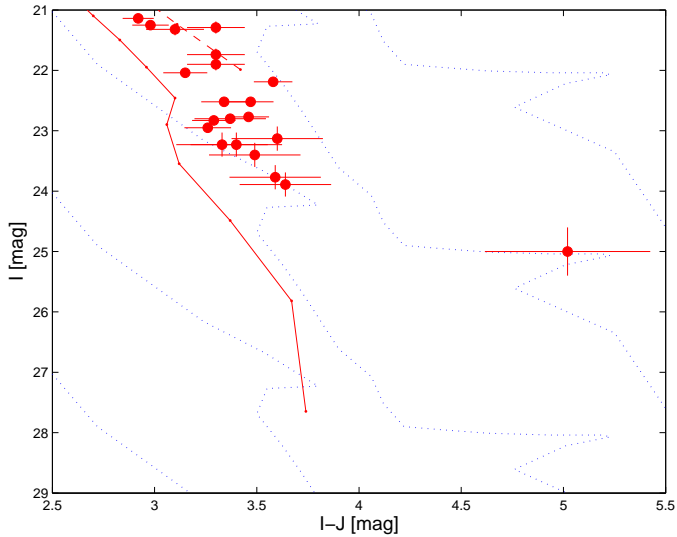}
\caption{$I$ vs. $I-J$ colour-magnitude diagram of the $\sigma$~Orionis
cluster.
Big filled circles with error bars: cluster members and candidates (the
red outlier is the T-type object S\,Ori~70);
solid and dashed lines: COND03 and DUSTY00 3\,Ma-old isochrones, respectively,
shifted at $d$ = 385\,pc (Chabrier et~al. 2000; Baraffe et~al. 2003);
dotted lines: dwarf sequence in Table~\ref{MLT} shifted at distance moduli
$m-M$ = 3.0, 6.0, 9.0, 12.0, and 15.0\,mag, from top to bottom.
The faintest data point of the COND03 theoretical isochrone ($I-J \approx$
3.0\,mag, $I \approx$ 29.0\,mag) corresponds to an 1.0\,$M_{\rm Jup}$-mass
object. 
L and T field dwarfs have approximate colours redder than $I-J \approx$ 3.5 and
4.5\,mag.}
\label{cmd}
\end{figure}

In short, from Table~\ref{contamination}, the number of ultracool
dwarfs in deep surveys is larger for earlier spectral types and for the
faintest magnitude intervals. 
The contamination by T dwarfs is very low for survey depths $I \lesssim$
24\,mag. 
Surveys deeper than this value are strongly affected by L- and T-type dwarf
interlopers.
From Fig.~\ref{cmd}, it is not clear what happens to the $\sigma$~Orionis
luminosity function at $I \gtrsim$ 24\,mag.
Currently, the deepest optical surveys in young clusters in general, and in
$\sigma$~Orionis in particular, reach limiting magnitudes $I$ = 24--25\,mag
(Caballero et~al. 2007 and references therein -- see also Comer\'on \&
Claes 2004 for a very deep near-infrared survey in Chamaeleon~I),
although some ultra-deep extragalactic surveys can go 2--4\,mag fainter. 
The magnitudes ($I \approx$ 22.8 and 23.2\,mag) and measured spectral types
(L5.0$\pm$1.0 and L3.5$\pm$2.0 -- Barrado y Navascu\'es et~al. 2001; Mart\'{\i}n
et a~al. 2001) of the faintest cluster objects with unambiguous youth features,
\object{S\,Ori~65} and \object{S\,Ori~60} (they display flux excess
at 8.0 and/or 5.8\,$\mu$m -- Zapatero Osorio et~al. 2007b; Scholz \&
Jayawardhana 2008) suggest  that the L-T transition in the cluster occurs at
about $I \sim$ 23.5\,mag (field) T0--1V dwarfs are about 1.5\,mag fainter
than L5--6V ones -- however, this assumption might not be necessarily right
for low-gravity sources due to pressure effects on alkali line absorption).  
\object{S\,Ori~69}, the second faintest cluster member {\em candidate}
($I \approx$ 23.9\,mag) has a tentative T0: spectral type from a
near-infrared spectrum (Mart\'{\i}n et~al. 2001), which supports our estimate
(the semi-colon indicates uncertainty in the classification). 
Therefore, fainter cluster members are expected to have T spectral types, 
with colours $I-J \gtrsim$ 4.5\,mag and most probable masses derived from
state-of-the-art theoretical models at $M \lesssim$ 0.006\,$M_\odot$ (we do not
discuss in this paper the cluster membership status of S\,Ori~70).

The actual number of $\sigma$~Orionis cluster members and candidates of L0--5
and L5--T0 spectral types at the corresponding magnitudes $I \sim$ 21.0--22.5
and 22.5--24.0\,mag in a survey is larger than the expected number of dwarf
contaminants of the same spectral type (e.g., Zapatero Osorio et~al. 2000). 
The difference, although significant, requires a careful
spectroscopic/photometric/astrometric follow-up because fluctuations about the
mean could mask the data. 
A different space density of contaminants towards $\sigma$~Orionis by some
Poissonian factor would easily lead the L-type contamination rate to grow
up from 36 to $\sim$70\,\% of more. 
Localized overdensities of dwarfs in the Galactic thin disc are not unusual
(e.g., Juri\'c et~al. 2008 and references therein). 
Field L dwarfs stop contaminating the $\sigma$~Orionis colour-magnitude diagram
at $I \approx$ 24.5\,mag (i.e., $I-J \gtrsim$ 4.5\,mag). 
Since the interstellar reddening is appreciable only at very large heliocentric
distances ($A_I = ad$, being $a \approx$ 2~10$^{-4}$\,mag\,pc$^{-1}$), where the
spatial density is extremely low, contamination by M dwarfs is not expected,
either.  

Surveys with completeness magnitudes $I_{\rm compl.} \approx$ 27--28\,mag {\em
should} be required to search for the opacity mass limit (at $M \lesssim$
0.003\,$M_\odot$), and field T dwarfs are expected to be the major
colour-selected contaminants at these magnitudes.  
While the contamination by T dwarfs in the brightest magnitude intervals in
Table~\ref{contamination} is very low with respect to the contamination by L
dwarfs (by a factor $\sim$100), it dramatically increases in the faintest
magnitude intervals.
For example, for $\Delta I$ = 28.0--29.0\,mag, the number of L0--L5V dwarf
contaminants reaches the maximum limit of $\sim$300 objects (i.e., the survey is
so deep that is able to detect {\em almost} all the early L dwarfs of the Galaxy
in that direction), whereas there are only 3.6 times less T dwarfs than L dwarfs
for the same $\Delta I$.
From the data in Table~\ref{contamination}, we expect about $\sim$45
T-type dwarfs in the foreground of $\sigma$~Orionis in an ultra-deep ($I
\approx$ 27--28\,mag) survey. 
However, the most optimistic estimations of a large index $\alpha$ = +0.6
of the cluster mass spectrum (Caballero et~al. 2007) extrapolated towards
only 0.001\,$M_\odot$ predict $\lesssim$30 planetary-mass objects with T
spectral types in $\sigma$~Orionis (Zapatero Osorio, priv.~comm.).
Assuming less favourable conditions, we forecast no more than $\sim$10 T-type
cluster members. 
These values depend on assumptions that require confirmation: e.g., the
distance and age of the cluster, the actual value of the slope of the mass
spectrum at Jovian masses, or the validity of theoretical evolutionary tracks
of low-mass stars and brown dwarfs at very young ages and low masses (Baraffe
et~al. 2002). 
For example, Fig.~\ref{cmd} shows that COND03 models (Baraffe et~al. 2003),
which are especially useful for the T-type domain in the field, predict rather
blue $I-J$ colours for the least massive hypothetical $\sigma$~Orionis members
(with masses down to $\sim$1.0\,$M_{\rm Jup}$).
Accurate theoretical modeling of low gravity, low temperature spectra is
still work in progress.

Even with the largest current ground facilities (2 $\times$ 10.0\,m Keck
Observatory, 10.4\,m Gran Telescopio Canarias), the necessary
follow-up of an ultra-deep survey for studying the opacity mass limit in
$\sigma$~Orionis would require a tremendous, prohibitive, observational effort.
For example, the low-resolution near-infrared spectrum of the $\sim$T6 object
S\,Ori~70 ($I$ = 25.0$\pm$0.4\,mag, $J$ = 19.98$\pm$0.06\,mag) needed a total
exposure time of 4\,800\,s with NIRSPEC at the Keck~II (Mart\'{\i}n \& Zapatero
Osorio 2003 -- it had, besides, a low signal-to-noise ratio).
Sources with $I \approx$ 27--28\,mag would need correspondingly larger exposure
times, of up to a whole night.
Therefore, spectroscopy of roughly half of the selected cluster member
candidates in such an ultra-deep survey ($\gtrsim$ 20 targets) would demand
about one month of observing time at the Keck~II. 
Shallower surveys ($I \lesssim$ 26\,mag) and their corresponding follow-ups
are, however, much less affected by T-dwarf contamination ($\sim$5 interlopers
in a 1\,deg$^2$-wide survey) and would demand a quite more reasonable amount of
time to be accomplished. 
The photometric selection becomes more stringent and may lower the
contamination level when combining several colour-magnitude diagrams as well as
colour-colour diagrams and CH$_4$-{\em on} and {\em off} photometry.
A survey with at least three broad filters (e.g., $IJK_{\rm s}$ or
$z'H[4.5]$)\footnote{$[4.5]$ is the 4.5\,$\mu$m IRAC passband at the {\em
Spitzer} Space Observatory, still useful during post-cryo phase.} and the two
narrow methane filters, although it multiplies the total exposure time by
a factor $\sim$5--6 with respect to an $IJ$ survey, may represent an
intermediate solution.

\section{Summary}

Late-type field dwarfs are the most important contributors to contamination
among faint, young, cluster member candidates selected from colour-magnitude
diagrams. 
The number of such interlopers in deep photometric surveys in clusters is of the
maximum  importance for the study of the substellar Initial Mass Function. 
We provide a pseudo-algorithm, an expression for the integrated number of
field dwarfs at certain heliocentric distance and galactic coordinates (assuming
an accurate linear approximation in the structure of the Galactic thin disc),
and the absolute magnitudes, colours and local spatial densities of M3--T8
dwarfs to compute the number of field late-type contaminants in deep surveys
excluding the Galactic~plane.

We have applied our tools and data to an hypothetical ultra-deep ($I_{\rm
compl.}$ = 29\,mag) survey towards the young ($\sim$3\,Ma) $\sigma$~Orionis
cluster.
We predict a rather low contamination rate of L and T field dwarfs with an
appropriately defined selection criterion up to $I$ = 25--26\,mag.
The number of contaminants at fainter magnitudes, where the opacity mass limit
is expected to lie, is however very large and may preclude photometric,
spectroscopic and astrometric follow-up in reasonable amounts of time with
current facilities. 

The enhancement of the contamination by T-type dwarfs in ultra- and very deep
surveys is not only important for $\sigma$~Orionis, in particular, and 
star-forming regions, in general, but also for extragalactic surveys of $z
\gtrsim$ 6 quasars, whose colours resemble those of T~dwarfs.

\begin{acknowledgements}

We thank the anonymous referee for his/her helpful report, I.~Baraffe for
providing us 3\,Ma-old Lyon tracks, and M.~Cornide and M.~R. Zapatero
Osorio for helpful comments.
J.A.C. formerly was an Alexander von Humboldt Fellow at the MPIA and currently
is an Investigador Juan de la Cierva at the UCM. 
R.K. is a granted PhD student at the MPIA.
Partial financial support was provided by the Universidad Complutense de Madrid
and the Spanish Ministerio Educaci\'on y Ciencia under grant 
AyA2005--02750 of the Programa Nacional de Astronom\'{\i}a y Astrof\'{\i}sica
and by the Comunidad Aut\'onoma de Madrid under PRICIT project S--0505/ESP--0237
(AstroCAM). 
Research has benefitted from the M, L, and T dwarf compendium housed at
{\tt DwarfArchives.org} and maintained by Chris Gelino, Davy Kirkpatrick, and
Adam Burgasser. 

\end{acknowledgements}

\end{document}